\documentclass[]{PoS}
\usepackage[authoryear,square]{natbib}
\bibpunct{(}{)}{;}{a}{}{,}

\makeindex
\citeindexfalse

\title{Observing Radio Pulsars in the Galactic Centre\\ with the Square Kilometre Array}

\ShortTitle{Galactic Centre Pulsars with the SKA}

\author{\speaker{R.~P.~Eatough}$^{1,2}$, T.~J.~W.~Lazio$^{3,4}$,
  J.~Casanellas$^5$, S.~Chatterjee$^6$, J.~M.~Cordes$^6$,
  P.~B.~Demorest$^7$, M.~Kramer$^{1,8}$, K.~J.~Lee$^{9,1}$, K.~Liu$^{1,10}$, S.~M.~Ransom$^7$, N.~Wex$^1$}

\author{\\\scriptsize
	$^1$Max-Planck-Institut f\"ur Radioastronomie, Bonn, 53121, Germany;
        $^2$E-mail: \email{reatough@mpifr-bonn.mpg.de};
        $^3$Jet Propulsion Laboratory, California Institute of Technology, M/S 138-308, 4800 Oak Grove Dr, Pasadena, CA 91109, USA;
        $^4$E-mail: \email{joseph.lazio@jpl.nasa.gov};
        $^5$Max Planck Institute for Gravitational Physics (Albert Einstein Institute), Science Park Potsdam-Golm, Am M\"uhlenberg 1, D-14476, Germany;
        $^6$Department of Astronomy and Space Sciences, Cornell University, Ithaca, NY 14853, USA;
        $^7$National Radio Astronomy Observatory, 520 Edgemont Road, Charlottesville, Virginia 22903-2475, USA;
        $^8$Jodrell Bank Centre for Astrophysics, Alan Turing Building, University of Manchester, Manchester, M13 9PL, UK;
        $^9$Kavli institute for Astronomy and Astrophysics, Peking University, Beijing 100871, P. R. China,;
        $^{10}$Station de radioastronomie de Nan\c{c}ay, Observatoire de Paris, CNRS/INSU, Universit\'{e} d'Orl\'{e}ans, F-18330 Nan\c{c}ay, France.
     }

\abstract{The discovery and timing of radio pulsars within the
  Galactic centre is a fundamental aspect of the SKA Science Case,
  responding to the topic of ``{\it Strong Field Tests of Gravity with
    Pulsars and Black Holes}'' (Kramer {et~al.}~2004; Cordes {et~al.}
  2004). Pulsars have in many ways proven to be excellent tools for
  testing the General theory of Relativity and alternative gravity
  theories (see Wex (2014) for a recent review). Timing a pulsar in
  orbit around a companion, provides a unique way of probing the
  relativistic dynamics and spacetime of such a system. The strictest
  tests of gravity, in strong field conditions, are expected to come
  from a pulsar orbiting a black hole. In this sense, a pulsar in a
  close orbit ($P_{\mathrm{orb}} < 1$~yr) around our nearest
  supermassive black hole candidate, Sagittarius~A* - at a distance of
  $\simeq8.3$~kpc in the Galactic centre (Gillessen {et~al.} 2009a) -
  would be the ideal tool. Given the size of the orbit and the
  relativistic effects associated with it, even a slowly spinning
  pulsar would allow the black hole spacetime to be explored in great
  detail (Liu {et~al.}  2012). For example, measurement of the frame
  dragging caused by the rotation of the supermassive black hole,
  would allow a test of the ``cosmic censorship conjecture.'' The
  ``no-hair theorem'' can be tested by measuring the quadrupole moment
  of the black hole. These are two of the prime examples for the
  fundamental studies of gravity one could do with a pulsar around
  Sagittarius~A*. As will be shown here, SKA1-MID and ultimately the
  SKA will provide the opportunity to begin to find and time the
  pulsars in this extreme environment.}

\FullConference{
Advancing Astrophysics with the Square Kilometre Array\\
June 8-13, 2014\\
Giardini Naxos, Italy}

\newcommand{\skipthis}[1]{}

\newcommand\arcsec{\hbox{$^{\prime\prime}$}}
\newcommand\arcmin{\hbox{$^\prime$}}
\newcommand\arcdeg{\hbox{$^\circ$}}

\newcommand{\sgra}{Sgr~A* }

\newcommand{\SM}{\rm SM}

\begin{document}
\section{Background}
\label{sec:PSR.GC.intro}

A combination of radio and infrared (IR) astrometric observations has
established the presence of a dark mass in the Galactic centre (GC),
whose properties are most consistent with being a supermassive black
hole (\hbox{BH}) \citep{eg96,Ghez98,rb04,Ghez+2008,gef+09,get+09}.
Known by its radio nomenclature of Sagittarius~A* (Sgr~A*), current
estimates are that the mass of this BH is $4.30 \pm
0.20_{\mathrm{(stat)}} \pm 0.30_{\mathrm{(sys)}} \times
10^6$~M${}_\odot$ \citep{gef+09}. Monitoring the orbits of the closest
stars S0-2 \citep{schoedel+02} and S0-102 \citep{meyer+12} could
provide constraints on the extent to which their orbits are perturbed
which in turn would constrain the enclosed mass, which could be stars,
dark matter, or both \citep{sabha+12,get+09}. Current work with the
Keck Telescope, the VLT and with the VLT interferometer (VLTI) and
proposed work with extremely large telescopes (ELTs; e.g. the European
ELT and the GSMT) will lead to IR astrometric precisions of 0.5~mas
with single apertures and 10~$\mu$as with the \hbox{VLTI} and the
GRAVITY instrument \citep{Weinberg+2005, eisenhauer+11}.  These
resolutions allow orbital perturbations of~4.15 and~0.08~au,
respectively, to be measured that will arise from General Relativity
(GR) and from Newtonian effects caused by other stars and dark matter.
It is expected that source confusion in IR surveys with ELTs will
limit the sample of stars to those with orbits larger than about~200
to~300~au; $\sim100$~stars with semi-major axes $\lesssim 3000$~au for
a 30-m ELT \citep{Weinberg+2005}.

The discovery and timing of radio pulsars orbiting Sgr~A*, and within
the central stellar cluster, will provide a powerful means to test GR,
as well as probe the dark matter content and magnetoionic medium
within the nucleus of our Galaxy.  An SKA radio pulsar program can
yield much more precise localizations on objects arbitrarily close to
the innermost stable circular orbit (\hbox{ISCO}, radius $\gtrsim
0.1$~au, depending on the spin of the Sgr~A* BH).  Confusion limits do
not apply to the time-varying signals of pulsars.  In comparison with
the best forecasted precisions of IR astrometry and Doppler
measurements, pulsar timing has the potential of increasing the
testing and diagnostic precision by {\em at least three orders of
  magnitude}.  Pulsars offer this prospect because timing precisions
correspond to spatial localizations (from {\em ranging}) smaller than
a light second (0.2~$\mu$as) compared to the localizations expected
from the best proposed IR interferometry, of no better than 10~$\mu$as
(0.1~au).  



\begin{figure}[tb]
\makebox[0.47\textwidth][c]{
	\hskip 0.6in 
	\includegraphics[width=0.58\textwidth]
		{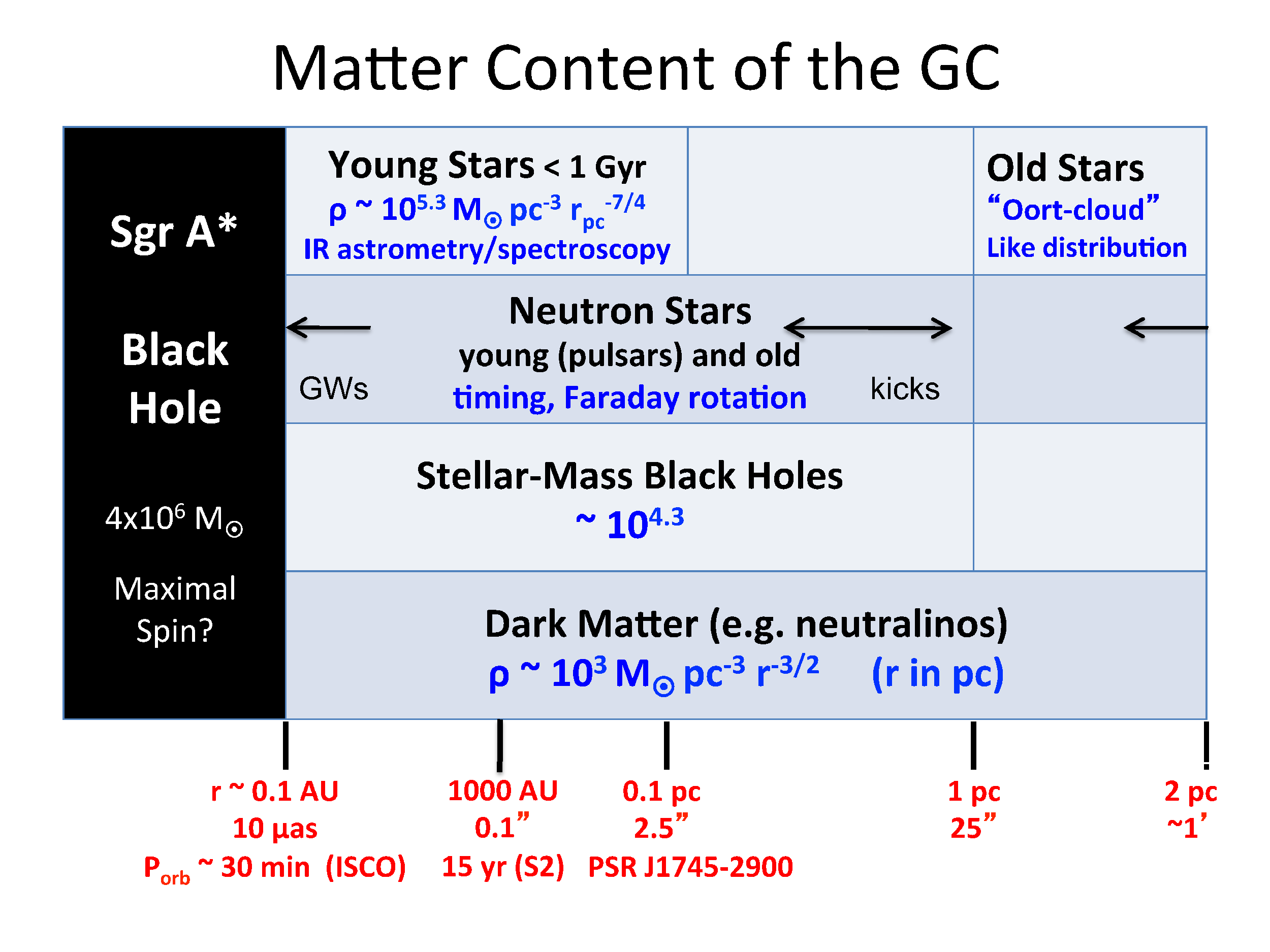}\\\vspace*{8ex}}\,\,%
	\hskip 0.6in 
        \includegraphics[width=0.41\textwidth]{atest2.ps}
\vspace{-0.7em}
\caption{ (\textit{Left}) Schematic view of the matter content of the
  \hbox{GC}.  Young stars (e.g., S0-2) are known to approach as close as
  0.1\arcsec\ (1000~AU) to Sgr~A*.  The combination of young, hot
  stars in the GC and mass segregation suggests that there are likely
  to be compact stellar remnants (NSs and BHs) at least
  within~\hbox{1000~au}, and potentially much closer.  Finally, there
  could also be dark matter particles.  The \hbox{SKA}, operating at
  frequencies near~10~GHz, will search for NS, in the form of active
  radio pulsars, as close as possible to Sgr~A*.  (\textit{Right})
  Depth of gravitational potential ($GM/ac^2$) probed by an orbit of
  semi-major axis~$a$ as a function of the mass of the binary
  system~$M$.  The regions probed by solar system tests and known
  pulsars in binary systems are indicated, with the double neutron
  star systems PSR~B1913$+$16 and PSR~J0737$-$3039 shown explicitly.
  The vertical cyan line indicates the mass of Sgr~A* (GC-BH).  The
  red line segment shows the region probed by the S stars in the GC.
  It is clear that substantially deeper gravitational potentials
  (larger $GM/ac^2$) remains to be probed, and would be accessible to
  radio pulsar observations \citep{SKAbook1a}.
\label{fig:GC.PSR.GCcontents}
}
\end{figure}

\section{The pulsar population in the GC}
\label{sec:PSR.GC.pops}

Several lines of argument suggest that not only is there a large
neutron star (NS) population in the \hbox{GC} but that many NSs should
be active radio pulsars.  Figure~\ref{fig:GC.PSR.GCcontents} (left
panel) shows schematically the contents of the GC as determined both
empirically and from theoretical arguments about galaxy formation.
\cite{Ghez+2008} and \cite{get+09} show that the stellar cluster around Sgr~A*
consists mainly of early-type stars, whose short lifetimes
($\sim10^7$~yr) means that either massive star formation in the region
is ongoing or that there has been a starburst of order 6~Myr ago,
although the presence of young, massive stars in a region where tidal
forces should inhibit star formation present a conundrum: ``the
paradox of youth'' - \citep{Ghez+2003}. Nonetheless, stars with masses
of~10--20~M${}_\odot$ are NS progenitors, and the presence of a
population of young massive stars near Sgr~A* suggests that there must
exist a numerous NS population as well. The time scales also work out
for many of these NSs to be active radio pulsars: radio emission of
canonical $10^{12}$~G objects lasts for $\sim 10$~Myr, the time
scale for spindown to lengthen spin periods past the ``death line'' of
a few seconds.

Estimates of the supernova rate within 100~pc of \sgra\ based on the
number of young stars in the region and on the heating of X-ray gas
range from~$10^{-3}$~~yr$^{-1}$ to~$10^{-2}$~yr$^{-1}$ and are
consistent, after accounting for the fraction of supernovae that
produce pulsars, with the number of point radio sources in the region
that could be pulsars \citep{lc08}. Such agreement is coarse but
suggestive that there are of order $10^3$~to~$10^4$ active pulsars in
the region, of which about~200 to~2000 (20\%) would be beamed toward
the Earth. More recently, and by combining results from the GC radio
point source counts described above, the detected GC pulsar population
at that time, non-detections in high-frequency pulsar surveys of the
inner parsec, radio and gamma-ray measurements of diffuse emission,
infra-red observations of massive stellar GC populations, pulsar wind
nebulae candidates, and the supernova rates from X-ray data,
\cite{wharton+12} estimate that $\sim10^3$ active radio pulsars in the
inner parsec should be beamed toward Earth.

Pulsars in the \hbox{GC}, like other stellar populations, will orbit
the central BH \sgra mostly as isolated objects but some will
be in binaries with companions of all types: main-sequence and
post-main-sequence stars, white dwarfs, \hbox{NSs}, or even \hbox{BHs}
\citep{fgal11}. A population of ``recycled'' pulsars (i.e., NSs spun up
to millisecond periods by accretion) is also expected because the
stellar density in the GC is large enough so that tidal capture and
exchange reactions should occur as they do in globular clusters.
\cite{vg07} show that tidal capture of NSs by low-mass main-sequence
stars is the dominant mechanism for forming binary NS in approximately
the central 1\arcmin\ of M31 and these can produce millisecond pulsars
(MSPs) via long-term accretion.  Along with similarly captured
stellar-mass BHs, accreting objects in the central part of M31 can
account for the excess number of point X-ray sources in the inner
bulge of M31.  The same processes should occur in the GC over a
similar-sized region (0.1--0.2~kpc) \citep{Muno+2005}.

\begin{figure}[bt]
\centering
\includegraphics[width=0.56\textwidth, angle=-90]{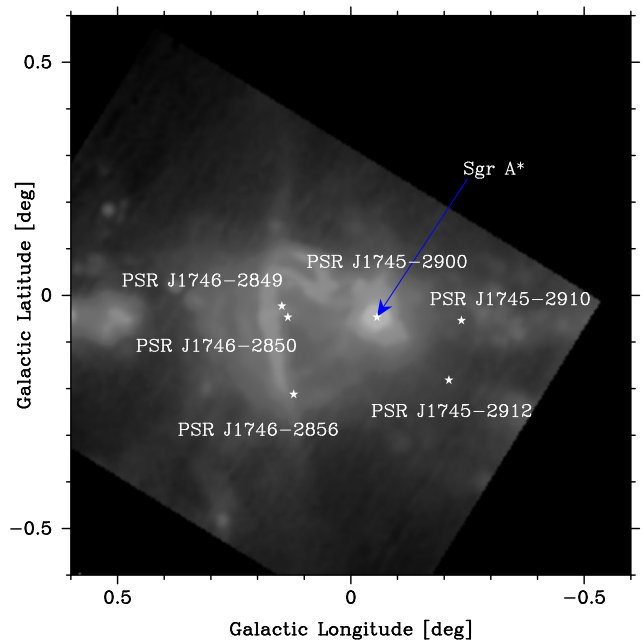}
\hspace{0.1in}
\caption{The positions of detected radio pulsars in the central
  0.5\arcdeg\ toward the GC overlaid on a 10.55~GHz continuum map made
  with the Effelsberg telescope \citep{seiradakis89}. Assuming a GC
  distance of $8.3\,{\rm kpc}$, 0.5\arcdeg corresponds to a projected
  distance of $\sim 70\,{\rm pc}$. Even PSR~J1745$-$2900, which
  overlaps the position of \sgra on this scale ($\sim3$\arcsec$\equiv
  0.1$~pc offset), is too distant for gravity tests. Figure courtesy
  of B.~Klein.
\label{fig:GC.PSR.gcpsrs}}
\end{figure}

Arguably the strongest evidence for a significant NS population in the
central parsec now comes from the recent detection of a radio loud
magnetar - PSR J1745$-$2900, an inherently rare class of pulsar, just
$\sim$3\arcsec (0.1~pc) from \sgra\ \citep{kennea+13, mori+13,
  eatough+13a, shannon+johnston+13}. Along with the five pulsars
previously discovered in proximity to \sgra (at 10\arcmin$-$15\arcmin
- see Figure~\ref{fig:GC.PSR.gcpsrs} \citep{j+06, dcl09}) these objects
cannot be explained as part of a Galactic disk population because the
number of foreground pulsars expected in the pulsar surveys to date is
much less than one. Consequently, the question of the GC pulsar
population has been recently revisited by a number of authors
\citep{chennamangalam+14, zhang+14, dexter+14}. As will be described in
Section~\ref{sec:PSR.GC.effects}, it is pulsars in compact orbits
($P_{\mathrm{orb}} < 1$~yr - corresponding to a semi-major axis of
$\sim~160$~au) around the BH that will enable the most precise tests
of GR. Recent simulations suggest that up $\sim200$ pulsars could
orbit within 4000~au of the BH, the closest of which could have a
semi-major axis of $\sim 120$~au \citep{zhang+14}. It is only through
observations with the SKA that the full GC pulsar population (whatever
its size) will be detected.

Certain dark matter models predict that dark matter particles can
efficiently accumulate inside compact objects such as NS, eventually
causing them to collapse into a BH. The GC pulsar population, as
detected by the SKA, can thus also be used to place stringent
constraints on dark matter models, with older pulsars and MSPs being
the most constraining \citep{bramante+14}.

\section{Testing gravity with GC pulsars}
\label{sec:PSR.GC.effects}

With their large moments of inertia and fast spin rates, radio pulsars
are excellent clocks. The clocklike property of pulsars orbiting
Sgr~A* will provide unique probes of the spacetime, plasma and dark
matter around the massive central BH and are a powerful probe of GR
\citep{pt79,wgs96,pl04} itself.  As we show below, GR and other
perturbations of GC pulsars are large enough so that even rather
``noisy'' pulsars, relative to typical precision timing requirements,
will do the job.

Pulse times-of-arrival (TOAs) from an orbiting pulsar can be described in
terms of the standard Keplerian parameters and in a model-independent
parameterized post-Newtonian (PPN) formalism.  Any deviations from the PPN values
expected from GR would provide strong evidence of a post-GR theory of
gravity.  The most recent illustration of this method of testing
theories of gravity is from the double pulsar
\hbox{PSR~J0737$-$3039A/B}, for which the measured PPN parameters
indicate agreement with GR at better than the 1\% level \citep{ks08}.
However, \hbox{PSR~J0737$-$3039A/B} still represents a relatively weak
field case.  Consider the depth of the gravitational potential to be
$GM/ac^2$, for a total system mass~$M$ and semi-major axis~$a$, with
$G$ and~$c$ being the usual physical constants.  The current
population of binary pulsars probes only to $GM/ac^2 \lesssim
10^{-5.5}$.  By contrast, a pulsar with an orbit comparable to the
star S0-2 ($\approx 15$~yr) would probe to $GM/ac^2 \sim 10^{-4.5}$;
pulsars with shorter orbital periods could probe to even larger values
of $GM/ac^2$ (Figure~\ref{fig:GC.PSR.GCcontents}).  Star-star interactions of
course can cause perturbations that make GR tests more difficult.

\begin{figure}[t!]
\centering
\includegraphics[width=0.860\textwidth]{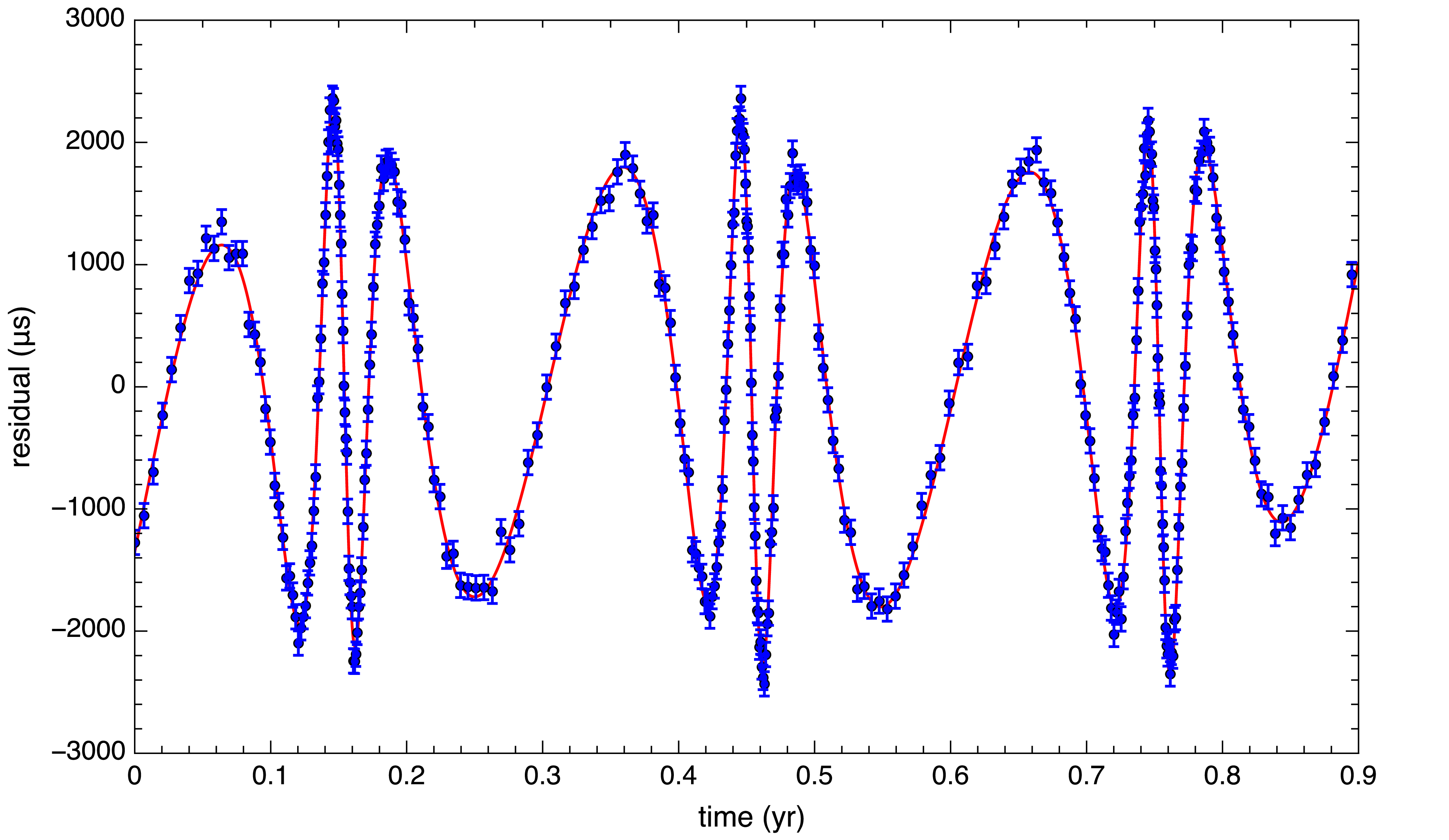}
\caption{Characteristic pulsar timing residuals caused by the
  quadrupole moment of the Sgr~A* BH in simulated times-of-arrival
  (TOAs) over three orbital cycles~\citep{liu+12}.  The orbital period
  is assumed to be $P_{\rm orb}=0.3\,{\rm yr}$ and the eccentricity
  $e=0.5$.  The red line is the fitted pulsar timing model.}
 \label{fig:residuals}
\end{figure}

\begin{figure}[tb]
\centering
\includegraphics[width=0.55\textwidth]{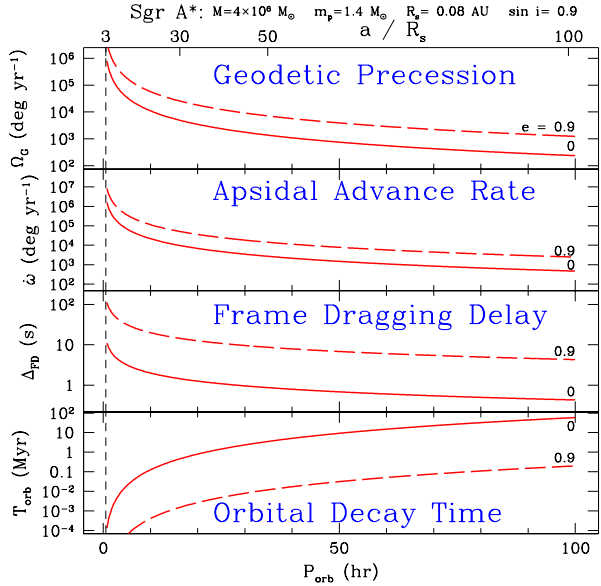}
\hskip 0.1in
\parbox[b]{0.35\textwidth}{%
 \caption{Relativistic effects that could be determined from measuring
   pulse TOAs from a pulsar closely orbiting Sgr~A*.  The
   \textit{bottom} axis shows the orbital period, and the \textit{top}
   axis shows the semi-major axis of the orbit, in units of the
   Schwarzschild radius of Sgr~A*.  All of the effects shown, except
   for frame dragging, have been detected in timing observations of
   binary pulsars.  In all panels, the solid line indicates a circular
   orbit ($e = 0$) while the dashed line indicates a highly eccentric
   orbit ($e = 0.9$).  For reference, the apsidal advance rates of
   PSR~J0737$-$3039, B1913$+$16, and Mercury are \emph{too small} to
   be plotted ($\dot\omega < 20\,\mathrm{deg}\,\mathrm{yr}^{-1}$).}
 \label{fig:GC.PSR.gr}
}
\end{figure}

For a pulsar in a close orbit ($P_{\mathrm{orb}} < 1$~yr) around
\sgra\ one expects to see a number of relativistic effects, like time
dilation, Shapiro delay and relativistic peri-center precession, all
of them with a magnitude, that as we mention above, does not even
require a particularly high timing precision to measure them with
excellent accuracy \citep{liu+12}.  The time dilation, for instance, is
expected to be of order 10 to 20 minutes. Consequently, soon after the
discovery relativistic effects will be well measured, give the mass of
Sgr A* with high precision, and allow for the first gravity
tests. After a few orbits, one will be able to separate the
Lense-Thirring drag from the orbital motion and measure the spin of
the BH with a fractional precision of $10^{-2}$ to $10^{-3}$
\citep{liu+12}. A first test is to see if the dimensionless
spin-parameter does not exceed one, as expected for a Kerr BH in GR
(cosmic censorship conjecture - see \cite{liu+12}). The quadrupole
moment of Sgr A* leads to a characteristic signature in the timing
residuals, directly related to the quadrupolar structure of its
potential (Figure~\ref{fig:residuals}). By this, the effect of the
quadrupole can uniquely be identified in the motion of the pulsar, and
its magnitude can be measured with good to very good precision,
allowing for a test of the no-hair theorem due to the unique relation
of mass, spin and quadrupole for a Kerr solution \citep{wk99,
  liu+12}. Figure~\ref{fig:GC.PSR.gr} illustrates some of the
relativistic effects that could be determined from measuring pulse
TOAs of a pulsar orbiting Sgr~A* in extremely short orbits
($P_{\mathrm{orb}} < 100$~hr).  For instance, measurement of the
relativistic apsidal advance, the second order Doppler shift, and the
Shapiro delay (not shown) yield redundant constraints on the
individual masses in the system as well as provide precision tests of
\hbox{GR}.  The Shapiro delay is detectable for elliptical orbits even
when the orbit is face on.  Geodetic precession of the spin axis will
cause secular or periodic changes in measured pulse shapes if there is
misalignment of the spin and orbital angular momenta, and secular
orbital decay from gravitational radiation will be measurable for the
more compact orbits ($P_{\mathrm{orb}} \lesssim 10$~hr).  Moreover, as
Figure~\ref{fig:GC.PSR.gr} demonstrates, only for these shortest
orbital periods will the orbital decay time become so small that we
will have little likelihood of detecting any such pulsar. Perhaps the
most notable aspect of GC pulsars is that orbital perturbations are
large enough so that even canonical ($10^{12}$~G) pulsars with long
periods ($\gtrsim 0.5$~s), which ordinarily are considered to be noisy
clocks, are suitable for making timing measurements.  This means that
{\em any} pulsar found in the GC is useful for the SKA pulsar program.

While many effects contribute to imprecision of TOAs \citep{cs10}, the
dominant ones for GC pulsars with canonical periods will be intrinsic
spin noise \citep{hlk10, sc10} and the large pulse widths that result
from scattering broadening.  With a conservative goal of about~1~s
timing ``precision'' (in quotes because detection of gravitational
waves with millisecond pulsars requires $< 100$~ns precision -
\cite{janssen+14}), many GR effects and orbital perturbations from
stellar encounters can be measured one to two orders of magnitude
better than the best IR interferometric measurements.  The RMS spin
noise for most objects will be below the 1~s target and the timing
error from the pulsar width~$W$ is $\delta t \sim W / {\rm (S/N)}$,
where S/N is the peak signal to noise ratio of the pulse.  For widths
$W < 1$~s, it is reasonable to expect $\delta t \sim 0.01$~to~0.1~s,
providing even more precise orbital measurements.

At present, all these considerations rest on the assumption that the
environment around Sgr A* is sufficiently clean, i.e. external
perturbations on the pulsar motion are negligible.  As shown in
Figure~\ref{fig:GC.PSR.GCcontents} the matter content of the GC
consists not only of \sgra, but early-type, late-type and evolved
stars, dark matter, and magnetized plasma all of which will cause many
perturbations that can be detected in pulsar TOAs. By measuring the
Newtonian precession of elliptical orbits, estimation of the smoothly
distributed mass, including both stars and dark matter, will be
possible \citep{Weinberg+2005}. In addition, strong constraints can be
made on the density of stars from two-body perturbations of pulsar
orbits \citep{Weinberg+2005, Merritt+2010}. To model potential external
perturbations to TOAs, and see how they can be accounted for, or how
much they degrade our capability to test GR, is subject to current
research.

Chromatic effects from intervening magnetized plasmas are discussed in
the next section; these effects are strongly frequency dependent and
can be removed (after extracting all valuable scientific information
on the properties of the GC interstellar medium) through appropriate
multi-frequency measurements.

\section{Chromatic radio propagation effects}
\label{sec:PSR.GC.prop}
Individual pulsar TOAs will show the well known dispersion and
multipath delays caused by the integrated electron column density (the
dispersion measure, DM) and by scattering from small-scale
($\lesssim$~1~au) density irregularities, characterized by the
scattering measure (SM).  Polarization measurements of pulse profiles
will show Faraday rotation from which the rotation measure (RM) can be
determined, thus constraining the integrated magnetic field.

Chromatic perturbations from magnetized plasma along the line of sight
include the differential {\em dispersion delay} across a narrow
bandwidth $\Delta\nu$ (in MHz), $\Delta t(\nu) =
8.3\,\mu\mathrm{s}\,\mathrm{DM}\,{\Delta\nu}/{\nu^3},$ where $\nu$ is
the observing frequency (in GHz) and DM is the \emph{dispersion
  measure} (in pc~cm${}^{-3}$), which is the electron column density,
$\mathrm{DM} \equiv \int dl\,n_e$.  Such delays are removed from
multi-frequency data both as part of the survey analysis (typically by
the use of \emph{filterbanks}) and also, with greater precision,
during timing measurements.

Towards the GC, a more important factor is pulse broadening from
scattering caused by fluctuations in electron density $\delta n_e$
that extend down to scales of a few hundred kilometers and have a
roughly Kolmogorov wavenumber spectrum \citep{ars95}. For the strong
scattering towards Sgr~A*, first identified by angular diameter
measurements of OH/IR masers and background AGNs, the pulse broadening
has a time scale $\tau_s \propto D \nu^{\sim -4}\SM$, where $D$ is the
pulsar distance and the scattering measure \SM\ is the line of sight
integral of $C_n^2$, the coefficient of the wavenumber spectrum for
$\delta n_e$ \citep{cl03}.  Unlike dispersion, which is easily removed
to maximize the S/N of a pulse, the deleterious effect of scatter
broadening cannot be corrected for when many
``scintles\footnote{Scintles are the constructive interference patches
  seen in the frequency-time plane for nearby pulsars. The scattering
  towards \sgra is strong enough to prohibit measurement of individual
  scintles because they are narrower than typical frequency channel
  widths even at higher radio frequencies e.g. 14~GHz.}'' are summed
across the observing bandwidth, as they will be for GC pulsars.

Magnetized plasma Faraday rotates the linear polarization by an angle
$\nu^{-2}\mathrm{RM}$ and is routinely measured for most pulsars using
full Stokes parameter pulse shape measurements. In the GC this effect
was demonstrated by measurements of the pulse polarization of the
recently discovered magnetar, PSR~J1745$-$2900, that revealed the
second largest RM known in the Galaxy (the largest being that of \sgra
itself) \citep{eatough+13a, shannon+johnston+13}. From X-ray
observations of the hot gas phase that \sgra is accreting from (and
which the magnetar is likely located in) a simple density profile was
used to place limits on the strength of the magnetic field in this
region; a value which can affect the dynamics of the accretion process
onto the supermassive BH \citep{eatough+13a}.

Once found, pulsars in the GC will therefore provide individual values
of all three measures (DM, \hbox{SM}, and RM) for different lines of
sight into the \hbox{GC}. Though not measured directly, the emission
measure EM (the integral of $n_e^2$) can be related to DM and SM so as
to constrain the ``outer scale'' of the wavenumber spectrum of $\delta
n_e$ \citep{cl02}.  The \hbox{EM}, while large toward \sgra, does not
cause significant free-free absorption that would prohibit detection
of pulsars as is clear since \sgra\ is seen through optically thin
plasma down to at least 1~GHz.

\section{Finding and timing pulsars in the GC with the SKA}
\label{sec:PSR.GC.surveys}

\subsection{Past searches and updates on GC scattering}
Despite predictions that a large number of radio loud pulsars should
be present in the GC (see Section 2.), at the time of writing just six
are known within 0.5\arcdeg\ (Figure~\ref{fig:GC.PSR.gcpsrs}). It has
been understood that the deficit in GC pulsars could be primarily
explained by extreme scattering of radio waves in the GC \citep{cl97,
  lc98, cl02, cl03}. As mentioned in Section 4. scattering, which
causes temporal broadening of pulses and a corresponding reduction in
pulse S/N, has a strong frequency dependence $(\propto \nu^{\sim
  -4})$.  A clear mitigation strategy for dealing with pulse
broadening due to scattering is to observe at frequencies much higher
than typically used for pulsar searching. The significant constraint
on this mitigation strategy is that typical pulsars have power-law
spectra ($S_\nu \propto \nu^\alpha$), usually with very steep spectra
indices ($\langle\alpha\rangle \simeq -1.7$) so their flux densities
decline rapidly with increasing frequency
\citep{kramer_spec98}. Fortunately, there is a subset of pulsars with
relatively flat spectra. For example, of~293 pulsars with derived
spectral indices in the ATNF
catalog\footnote{http://www.atnf.csiro.au/people/pulsar/psrcat/}
\citep{mhth05}, approximately 40 (or 14\%) have $\alpha > -1$.
Moreover, a small number of pulsars have been detected at frequencies
as high as 90~GHz \citep{kxj+97,mkt+97}. With such pulsars in mind,
surveys of the GC at progressively higher radio frequencies have been
performed \citep{klein04, j+06, dcl09, deneva-phdth, macquart+10,
  bates+11, eatough+13b, siemion+13} and resulted in the detection of
five of the six pulsars displayed in Figure~\ref{fig:GC.PSR.gcpsrs}
\citep{j+06, dcl09}; albeit at lower frequencies $\lesssim 3$~GHz.

The sixth known GC pulsar, PSR J1745$-$2900, was first identified at
X-ray wavelengths \citep{kennea+13, mori+13} before detection in the
radio \citep{eatough+13a, shannon+johnston+13}. Recent measurements of
the pulse broadening in this pulsar indicate the degree of scattering
toward the GC is less than previously predicted \citep{spitler14}. In
addition, measurements with the Very Long Baseline Array (VLBA) of the
angular scatter broadening show that it has an angular size equivalent
to \sgra itself. This suggests that PSR J1745$-$2900 and \sgra are
behind the same scattering region. This fact coupled with the smaller
pulse broadening time indicate that the scattering region (if modelled
as a single thin screen of material) could be distant from the GC
\citep{bower14}. However, the authors note that the paucity of GC
pulsars still indicates that regions of intense scattering might exist
in the GC. These regions could be of a more complex structure than in
the simple picture described above \citep{spitler14}. Therefore high
observing frequencies ($\gtrsim 9$~GHz) might still be necessary to
detect GC pulsars.

In the following section we use the latest information about GC
scattering to re-investigate the prospects for GC pulsar searches with
the SKA.

\subsection{Prospects for GC pulsar searches with the SKA}
\begin{figure}[tb]
\centering \includegraphics[width=0.7\textwidth]{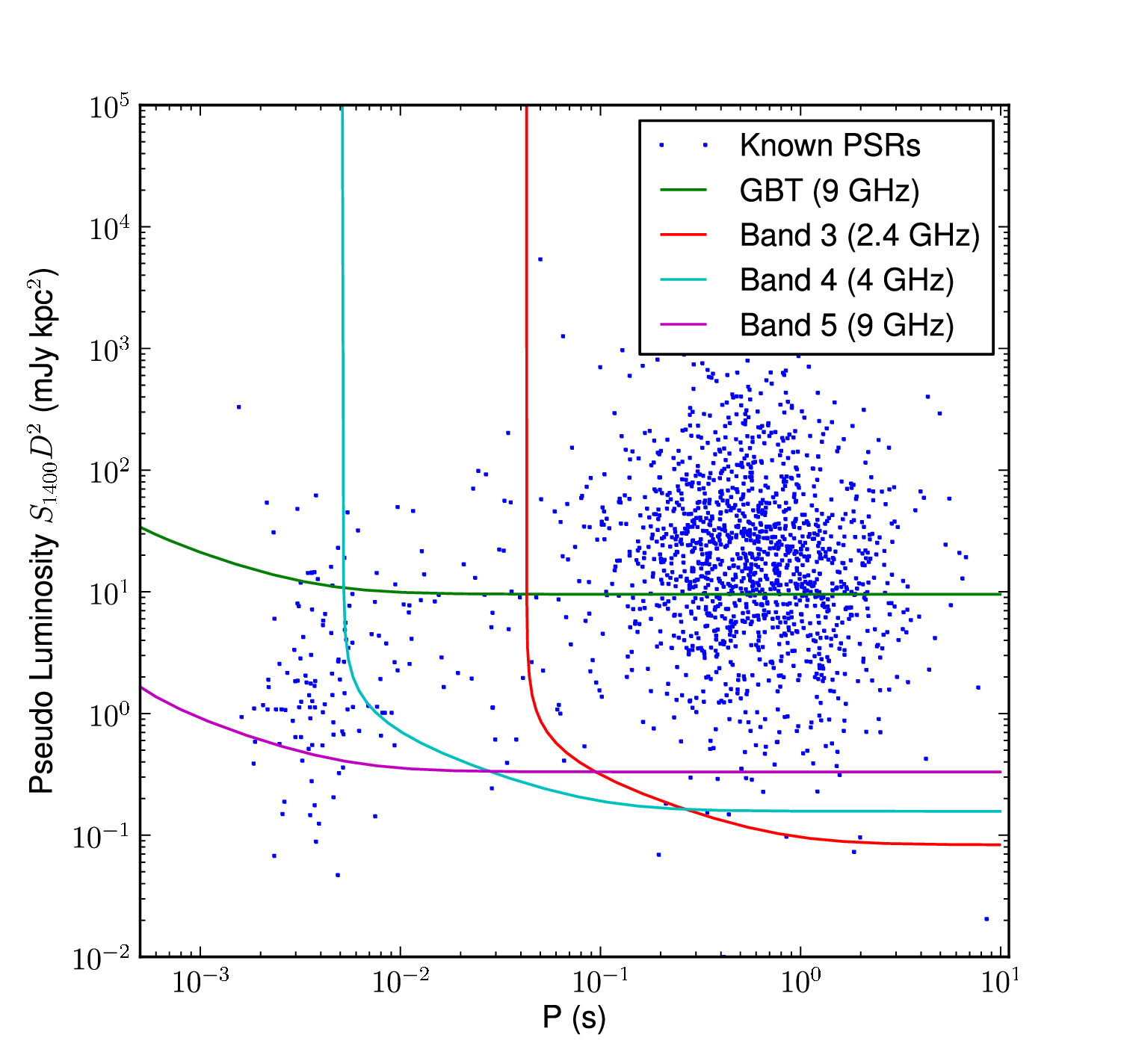}
\vspace*{-1ex}
\caption{Detectability of pulsars in the GC via a periodicity search
  of a 6~hr integration with the GBT at 9~GHz and SKA1-MID at three of the
  proposed frequency bands. Points denote the period-averaged ``pseudo
  luminosity'' at~1.4~GHz ($L = S_{1400} D^2$) versus spin period for
  known pulsars in the ATNF catalog \citep{mhth05}. The curves show
  sensitivity limits derived from the minimum detectable flux density
  (Equation~5.1). Pulsars are detectable if they are \emph{above} a
  given curve. We have used a GC distance of~8.3~kpc; a S/N\,$=10$
  detection threshold; a bandwidth $\Delta \nu=$~0.8~GHz for the GBT,
  and bandwidths $\Delta \nu =$~1.4, 2.4 and 9.2~GHz for SKA1-MID Bands 3,
  4 and 5 respectively \citep{baseline_design}; a system temperature,
  neglecting the effects background sky temperature, of $T_{\rm
    sys}=$27~K; and a spectral index of $\alpha = -1.7$, while
  recognizing that some objects have flatter and some steeper
  spectra. Pulse broadening, $\tau_{s}$, values in the direction of
  Sgr~A* are provided by measurements from the GG magnetar
  PSR~J1745$-$2900 \citep{spitler14}. The curves will move lower for
  pulsars having spectra less steep than $\nu^{-1.7}$. The curves will
  also move lower if multiple integrations are combined (e.g., through
  incoherent combination of power spectra \citep{eatough+13b}). On the
  other hand, if for example the sensitivity of the core of SKA1-MID that
  is used for the pulsar search is reduced, by making the collecting
  area two thirds of the proposed size, the curves will move upward by
  an amount of log(1.5).}
\label{fig:GC.PSR.lumvsp}
\end{figure}

Our top goal is to find and then time radio pulsars within the central
parsec (30\arcsec\ radius), with the highest priority placed upon
pulsars as close as possible to \sgra.  A secondary goal is to conduct
a more complete census of the larger region ($\sim 30\arcmin$) in
order to understand better the overall GC NS population.  While the
pulse broadening time due to scattering appears to be less than
previously predicted, the effect is still large enough to necessitate
pulsar searches of the GC at higher frequencies. In
Figure~\ref{fig:GC.PSR.lumvsp} we show the detection sensitivity of
pulsar periodicity searches of the GC performed with SKA1-MID and
utilizing three of the proposed frequency bands outlined in
\citep{baseline_design}. The figure shows the radio ``pseudo
luminosity'' at 1.4~GHz (given by $L_{\rm p}=S_{\rm 1400}D^2$, where
$S_{\rm 1400}$ and $D$ are the period-averaged flux density at 1.4~GHz
and the distance respectively) versus the spin period. The dots are
the actual values for known pulsars in the current (May 2014) ATNF
catalog \citep{mhth05}. The curves plotted for each frequency band are
derived from estimations of the minimum detectable flux density,
$S_{\mathrm{min}}$, given by the radiometer equation as applied to
pulsar observations,
\begin{equation}
S_{\mathrm{min}} = \beta \frac{{\rm S/N_{\rm min}}T_{\rm sys}}{G\sqrt{N_{\rm pol}~\Delta\nu~T_{\rm obs}}}\sqrt{\frac{W_{\rm eff}}{P-W_{\rm eff}}},
\label{eqn:smin}
\end{equation}
where $\beta$ is a digitization factor, ${\rm S/N_{\rm min}}$ is the
detection threshold, $T_{\rm sys}$ is the system temperature, $G$ is
the telescope gain (in units of ${\rm K}\,{\rm Jy}^{-1}$), $N_{\rm
  pol}$ is the number of polarization channels summed, $\Delta\nu$ is
the bandwidth, $T_{\rm obs}$ is the observation time, $P$ is the
pulsar spin period and $W_{\rm eff}$ is the effective pulse width.
Certain parameters have fairly standard values, namely, $\beta \approx
1$, $N_{\rm pol} = 2$, and ${\rm S/N_{\rm min}\approx 10}$. The
effective pulse width is given by the vector sum of the intrinsic
pulse width and any unaccounted for pulse broadening effects in the
following:
\begin{equation}
W_{\rm eff} = \sqrt{(W_{\mathrm{50}}^2 +\tau_{\rm samp}^2 +\tau_{\rm DM}^2 +\tau_{\rm s}^2)}
\end{equation}
where $W_{\mathrm{50}}$ is the intrinsic pulse width at half the pulse
height (here assumed to be $0.05P$), $\tau_{\rm samp}$ is the data
sampling interval, $\tau_{\rm DM}$ is the intra-channel dispersion
broadening (in a filterbank) and $\tau_{\rm s}$ is the scatter
broadening. The curves, which mark the minimum detectable pulsar spin
period and luminosity, are plotted for pulsars at the distance of the
GC and take into account the effects of scatter broadening, based on
measurements of $\tau_{\rm s}$ from PSR~J1745$-$2900 \citep{spitler14},
and assume a typical spectral index of $\alpha = -1.7$, thereby
allowing the limits to be scaled to the equivalent sensitivity at
1.4~GHz. Further details of the numbers used in defining these curves
are given in the caption to Figure~\ref{fig:GC.PSR.lumvsp}.

If the scattering toward PSR~J1745$-$2900 is representative of the
inner GC as a whole Figure~\ref{fig:GC.PSR.lumvsp} shows that the
common pulsars, with spin periods around 0.5~s, might be detectable at
frequencies as low as 2.4~GHz (Band 3) with SKA1-MID. For mildly recycled
pulsars, with spin periods from a few milliseconds to a few tens of
milliseconds, 4~GHz (Band 4) observations would be needed. Frequencies
up to at least 9~GHz (Band 5) might be required to detect the longer
lived MSPs in the GC. As well as the possibility of detecting MSPs, GC
pulsar surveys at higher frequencies offer the additional benefit of
reduced background sky temperatures, improving sensitivity.

For comparison with current facilities a line for a 9~GHz GC search
with the Green Bank Telescope (GBT) has also been marked. The fact
that a large fraction of the known pulsar population is already above
this line might be of cause for concern for the prospects for finding
more GC pulsars, however, \citep{chennamangalam+14} note that current
non-detections from past GC surveys cannot yet rule out a large GC
pulsar population which still need deeper surveys to be detected. SKA1-MID
and the SKA are needed to address this problem.

Our search methodology consists of both {\em periodicity searches},
including correction for orbital motion, and {\em searches for
  transient sources}, because some pulsars are more easily detectable
in single-pulse detection algorithms than in Fourier searches.  In
addition, the potential for discovering new classes of objects is
maximized by searching for individual bursts.

\begin{description}
\item[Periodicity Searches:]%
Periodicity searches are a standard tool for searching for pulsars
and have been used for many years, in many directions, not limited to the \hbox{GC}.  These algorithms
include standard de-dispersion to remove dispersion delays, orbital
demodulation, Fourier analysis of the resulting time series, and
identification of signals above threshold, which requires
classification as interference, instrumental contamination, or 
pulsars.  For the \hbox{GC}, the most important aspect of algorithmic
development concerns orbital demodulation, as it is for the shortest
orbital period pulsars that the maximum GR effects can be measured.
The simplest orbital demodulation is an acceleration search, which
removes a parabolic signature as an approximation to a Keplerian orbit.
This approach works for data sets that have a duration which is a small 
fraction of the orbital period \citep{rce+03}.  

\item[Transient Searches:]%
For compact orbits around \sgra, such as those $< 100$~hr shown in
Figure~\ref{fig:GC.PSR.gr}, the large GR effects that make such pulsars
extremely attractive targets will also make them more difficult to detect.
Fast geodetic precession, for example, may mimic an orbital period
over a 6~hr observation time and may be detectable at the output of
an acceleration search. However, precession of the pulsar beam will
make objects intermittent because the beam will precess
out of our line of sight for a sizable fraction of the precession
period.  This effect has been seen in several of the NS-NS binaries,
including the Hulse-Taylor pulsar \citep{kramer_geodetic98}.   Consequently, single epoch search
observations may be inadequate to find such sources.
Moreover, some pulsars emit ``giant'' pulses, with strengths that are
$1000\times$ the mean pulse intensity \citep{hr75,hkwe03,cstt96,jr03}.
In the case of the Crab pulsar, some giant pulses outshine the entire
Crab Nebula \citep{hr75}, corresponding to brightness temperatures of
order $10^{31}$~\hbox{K}; \citep{hkwe03} observed ``nano-giant'' pulses
having durations of only 2~ns and implied brightness temperatures
of~\hbox{$10^{38}$~K}, by far the most luminous emission from any
astronomical object.
\end{description}

Once GC pulsars are found with the SKA, our approach will be similar
to that already used by e.g. \citeauthor{dcl09}~(2009), namely they will be timed
over short and moderate time periods to characterize their spin
properties and any orbital motion.  Each will be observed at multiple
frequencies to find the optimal band for timing precision, which
depends on period, pulse width, flux density and spectral index, and
on the exact level of scattering.  Depending on the object, we will
then embark on a long-term timing campaign for measuring GR effects
and probing the GC environment, with a cadence determined by the
orbital period around \sgra\ and on any stellar orbital motion.

\end{document}